\documentclass[12pt,a4paper]{article}

\usepackage{ifthen} 
\newboolean{pdflatex}
\setboolean{pdflatex}{true} 

\newboolean{articletitles}
\setboolean{articletitles}{true} 

\newboolean{uprightparticles}
\setboolean{uprightparticles}{false} 

\newboolean{inbibliography}
\setboolean{inbibliography}{false} 

\usepackage[top=1in, bottom=1.25in, left=1in, right=1in]{geometry}

\usepackage{microtype}
\usepackage{lineno}  
\usepackage{xspace} 
\usepackage{caption} 

\usepackage{graphicx}  
\usepackage{color}
\usepackage{colortbl}
\graphicspath{{./figs/}} 

\usepackage{amsmath} 
\usepackage{amssymb}
\usepackage{amsfonts}
\usepackage{upgreek} 

\newcommand*\patchAmsMathEnvironmentForLineno[1]{%
\expandafter\let\csname old#1\expandafter\endcsname\csname #1\endcsname
\expandafter\let\csname oldend#1\expandafter\endcsname\csname
end#1\endcsname
 \renewenvironment{#1}%
   {\linenomath\csname old#1\endcsname}%
   {\csname oldend#1\endcsname\endlinenomath}%
}
\newcommand*\patchBothAmsMathEnvironmentsForLineno[1]{%
  \patchAmsMathEnvironmentForLineno{#1}%
  \patchAmsMathEnvironmentForLineno{#1*}%
}
\AtBeginDocument{%
\patchBothAmsMathEnvironmentsForLineno{equation}%
\patchBothAmsMathEnvironmentsForLineno{align}%
\patchBothAmsMathEnvironmentsForLineno{flalign}%
\patchBothAmsMathEnvironmentsForLineno{alignat}%
\patchBothAmsMathEnvironmentsForLineno{gather}%
\patchBothAmsMathEnvironmentsForLineno{multline}%
\patchBothAmsMathEnvironmentsForLineno{eqnarray}%
}

\usepackage{hyperref}    
\usepackage[all]{hypcap} 

\usepackage{xspace} 
\usepackage{upgreek}

\def\lhcb {\mbox{LHCb}\xspace}

\def\lhc    {\mbox{LHC}\xspace}

\def\rich   {RICH\xspace}

\def\MagUp {\mbox{\em Mag\kern -0.05em Up}\xspace}

\ifthenelse{\boolean{uprightparticles}}%
{

 \def\Pmu         {\ensuremath{\upmu}\xspace}

 \def\Ppi         {\ensuremath{\uppi}\xspace}

 \def\Pphi        {\ensuremath{\upphi}\xspace}

 \def\Ppsi        {\ensuremath{\uppsi}\xspace}

 \def\PDelta      {\ensuremath{\Delta}\xspace}                 
 \def\PXi      {\ensuremath{\Xi}\xspace}                 
 \def\PLambda      {\ensuremath{\Lambda}\xspace}                 
 \def\PSigma      {\ensuremath{\Sigma}\xspace}                 
 \def\POmega      {\ensuremath{\Omega}\xspace}                 
 \def\PUpsilon      {\ensuremath{\Upsilon}\xspace}                 
 
 \def\PB      {\ensuremath{\mathrm{B}}\xspace}                 
                  
 \def\PD      {\ensuremath{\mathrm{D}}\xspace}

 \def\PJ      {\ensuremath{\mathrm{J}}\xspace}                 
 \def\PK      {\ensuremath{\mathrm{K}}\xspace}

 \def\Pb      {\ensuremath{\mathrm{b}}\xspace}                 
 \def\Pc      {\ensuremath{\mathrm{c}}\xspace}                 
                  
 \def\Pe      {\ensuremath{\mathrm{e}}\xspace}

 \def\Pi      {\ensuremath{\mathrm{i}}\xspace}

 \def\Pp      {\ensuremath{\mathrm{p}}\xspace}

 \def\Ps      {\ensuremath{\mathrm{s}}\xspace}

}
{

 \def\Pmu         {\ensuremath{\mu}\xspace}

 \def\Ppi         {\ensuremath{\pi}\xspace}

 \def\Pphi        {\ensuremath{\phi}\xspace}

 \def\Ppsi        {\ensuremath{\psi}\xspace}                 
                  
 \mathchardef\PDelta="7101
 \mathchardef\PXi="7104
 \mathchardef\PLambda="7103
 \mathchardef\PSigma="7106
 \mathchardef\POmega="710A
 \mathchardef\PUpsilon="7107
                  
 \def\PB      {\ensuremath{B}\xspace}                 
                  
 \def\PD      {\ensuremath{D}\xspace}

 \def\PJ      {\ensuremath{J}\xspace}                 
 \def\PK      {\ensuremath{K}\xspace}

 \def\Pb      {\ensuremath{b}\xspace}                 
 \def\Pc      {\ensuremath{c}\xspace}                 
                  
 \def\Pe      {\ensuremath{e}\xspace}

 \def\Pi      {\ensuremath{i}\xspace}

 \def\Pp      {\ensuremath{p}\xspace}

 \def\Ps      {\ensuremath{s}\xspace}

}

\makeatletter
\ifcase \@ptsize \relax
  \newcommand{\miniscule}{\@setfontsize\miniscule{4}{5}}
\or
  \newcommand{\miniscule}{\@setfontsize\miniscule{5}{6}}
\or
  \newcommand{\miniscule}{\@setfontsize\miniscule{5}{6}}
\fi
\makeatother

\DeclareRobustCommand{\optbar}[1]{\shortstack{{\miniscule (\rule[.5ex]{1.25em}{.18mm})}
  \\ [-.7ex] $#1$}}

\def\epm        {{\ensuremath{\Pe^\pm}}\xspace} 
\def\epem       {{\ensuremath{\Pe^+\Pe^-}}\xspace}

\def\muon       {{\ensuremath{\Pmu}}\xspace}
\def\mup        {{\ensuremath{\Pmu^+}}\xspace}
\def\mumu       {{\ensuremath{\Pmu^+\Pmu^-}}\xspace}

\def\squark    {{\ensuremath{\Ps}}\xspace}

\def\cquark    {{\ensuremath{\Pc}}\xspace}

\def\bquark    {{\ensuremath{\Pb}}\xspace}

\def\pion   {{\ensuremath{\Ppi}}\xspace}

\def\pip    {{\ensuremath{\pion^+}}\xspace}
\def\pim    {{\ensuremath{\pion^-}}\xspace}
\def\pipm   {{\ensuremath{\pion^\pm}}\xspace}

\def\kaon    {{\ensuremath{\PK}}\xspace}
  \def\Kbar    {{\kern 0.2em\overline{\kern -0.2em \PK}{}}\xspace}

\def\KorKbar    {\kern 0.18em\optbar{\kern -0.18em K}{}\xspace}

\def\Kp      {{\ensuremath{\kaon^+}}\xspace}
\def\Km      {{\ensuremath{\kaon^-}}\xspace}
\def\Kpm     {{\ensuremath{\kaon^\pm}}\xspace}

\def\KS      {{\ensuremath{\kaon^0_{\mathrm{ \scriptscriptstyle S}}}}\xspace}

  \def\Dbar    {{\kern 0.2em\overline{\kern -0.2em \PD}{}}\xspace}
\def\D       {{\ensuremath{\PD}}\xspace}

\def\DorDbar    {\kern 0.18em\optbar{\kern -0.18em D}{}\xspace}
\def\Dz      {{\ensuremath{\D^0}}\xspace}

\def\Dstarp  {{\ensuremath{\D^{*+}}}\xspace}

\def\Dsp     {{\ensuremath{\D^+_\squark}}\xspace}

\def\B       {{\ensuremath{\PB}}\xspace}
\def\Bbar    {{\ensuremath{\kern 0.18em\overline{\kern -0.18em \PB}{}}}\xspace}

\def\BorBbar    {\kern 0.18em\optbar{\kern -0.18em B}{}\xspace}

\def\Bu      {{\ensuremath{\B^+}}\xspace}

\def\Bp      {{\ensuremath{\Bu}}\xspace}

\def\jpsi     {{\ensuremath{{\PJ\mskip -3mu/\mskip -2mu\Ppsi\mskip 2mu}}}\xspace}

  \def\Y#1S{\ensuremath{\PUpsilon{(#1S)}}\xspace}

\def\proton      {{\ensuremath{\Pp}}\xspace}
\def\antiproton  {{\ensuremath{\overline \proton}}\xspace}

\def\Lz          {{\ensuremath{\PLambda}}\xspace}
\def\Lbar        {{\ensuremath{\kern 0.1em\overline{\kern -0.1em\PLambda}}}\xspace}
\def\LorLbar    {\kern 0.18em\optbar{\kern -0.18em \PLambda}{}\xspace}

\def\Lc      {{\ensuremath{\Lz^+_\cquark}}\xspace}

\newcommand{\decay}[2]{\ensuremath{#1\!\to #2}\xspace}         

\def\to                 {\ensuremath{\rightarrow}\xspace}

\newcommand{\eff}{\ensuremath{\varepsilon}\xspace}
\newcommand{\avgeff}{\ensuremath{\bar{\varepsilon}}\xspace}

\def\AT#1     {\ensuremath{A_{\mathrm{T}}^{#1}}\xspace}           

\def\C#1      {\ensuremath{\mathcal{C}_{#1}}\xspace}                       
\def\Cp#1     {\ensuremath{\mathcal{C}_{#1}^{'}}\xspace}                    
\def\Ceff#1   {\ensuremath{\mathcal{C}_{#1}^{\mathrm{(eff)}}}\xspace}        
\def\Cpeff#1  {\ensuremath{\mathcal{C}_{#1}^{'\mathrm{(eff)}}}\xspace}       
\def\Ope#1    {\ensuremath{\mathcal{O}_{#1}}\xspace}                       
\def\Opep#1   {\ensuremath{\mathcal{O}_{#1}^{'}}\xspace}                    

\newcommand{\tev}{\ifthenelse{\boolean{inbibliography}}{\ensuremath{~T\kern -0.05em eV}\xspace}{\ensuremath{\mathrm{\,Te\kern -0.1em V}}}\xspace}
\newcommand{\gev}{\ensuremath{\mathrm{\,Ge\kern -0.1em V}}\xspace}
\newcommand{\mev}{\ensuremath{\mathrm{\,Me\kern -0.1em V}}\xspace}
\newcommand{\kev}{\ensuremath{\mathrm{\,ke\kern -0.1em V}}\xspace}
\newcommand{\ev}{\ensuremath{\mathrm{\,e\kern -0.1em V}}\xspace}
\newcommand{\gevc}{\ensuremath{{\mathrm{\,Ge\kern -0.1em V\!/}c}}\xspace}
\newcommand{\mevc}{\ensuremath{{\mathrm{\,Me\kern -0.1em V\!/}c}}\xspace}
\newcommand{\gevcc}{\ensuremath{{\mathrm{\,Ge\kern -0.1em V\!/}c^2}}\xspace}
\newcommand{\gevgevcccc}{\ensuremath{{\mathrm{\,Ge\kern -0.1em V^2\!/}c^4}}\xspace}
\newcommand{\mevcc}{\ensuremath{{\mathrm{\,Me\kern -0.1em V\!/}c^2}}\xspace}

\def\mum  {\ensuremath{{\,\upmu\mathrm{m}}}\xspace}

\def\invfb   {\ensuremath{\mbox{\,fb}^{-1}}\xspace}

\def\mhz  {\ensuremath{{\mathrm{ \,MHz}}}\xspace}

\def\ith{\ensuremath{i\text{th}}\xspace}

\def\gsim{{~\raise.15em\hbox{$>$}\kern-.85em
          \lower.35em\hbox{$\sim$}~}\xspace}
\def\lsim{{~\raise.15em\hbox{$<$}\kern-.85em
          \lower.35em\hbox{$\sim$}~}\xspace}

\def\sPlot{\mbox{\em sPlot}\xspace}

\def\ptot       {\mbox{$p$}\xspace}
\def\pt         {\mbox{$p_{\mathrm{ T}}$}\xspace}

\def\isMuon     {\ensuremath{\mathtt{isMuon}}\xspace}

\def\tell1  {TELL1\xspace}
\def\ukl1   {UKL1\xspace}

\newcommand{\ie}{\mbox{\itshape i.e.}\xspace}

\usepackage{cite} 
\usepackage{mciteplus}

\usepackage{longtable} 
\usepackage{multirow}

\def\sPlot{\ensuremath{_s\mathcal P\mathrm{lot}}\xspace}
\def\sWeight{\ensuremath{_s\mathcal W\mathrm{eight}}\xspace}
\def\sWeights{\ensuremath{_s\mathcal W\mathrm{eights}}\xspace}

\def\Full{\texttt{Full}\xspace}

\def\Turbo{\texttt{Turbo}\xspace}
\def\TurboCalib{\texttt{TurboCalib}\xspace}

\def\calo   {\mbox{CALO}\xspace}

\def\runo   {\mbox{Run 1}\xspace}
\def\runt   {\mbox{Run 2}\xspace}

\def\Lzz    {{\ensuremath{\Lz^0}}\xspace}

\def\prp    {{\ensuremath{\proton}}\xspace}

\def\mupm   {{\ensuremath{\muon^\pm}}\xspace}

\begin{document}

\renewcommand{\thefootnote}{\fnsymbol{footnote}}
\setcounter{footnote}{1}


\begin{titlepage}
\pagenumbering{roman}

\vspace*{-1.5cm}
\centerline{\large EUROPEAN ORGANIZATION FOR NUCLEAR RESEARCH (CERN)}
\vspace*{1.5cm}
\noindent
\begin{tabular*}{\linewidth}{lc@{\extracolsep{\fill}}r@{\extracolsep{0pt}}}
\ifthenelse{\boolean{pdflatex}}
{\vspace*{-2.7cm}\mbox{\!\!\!\includegraphics[width=.14\textwidth]{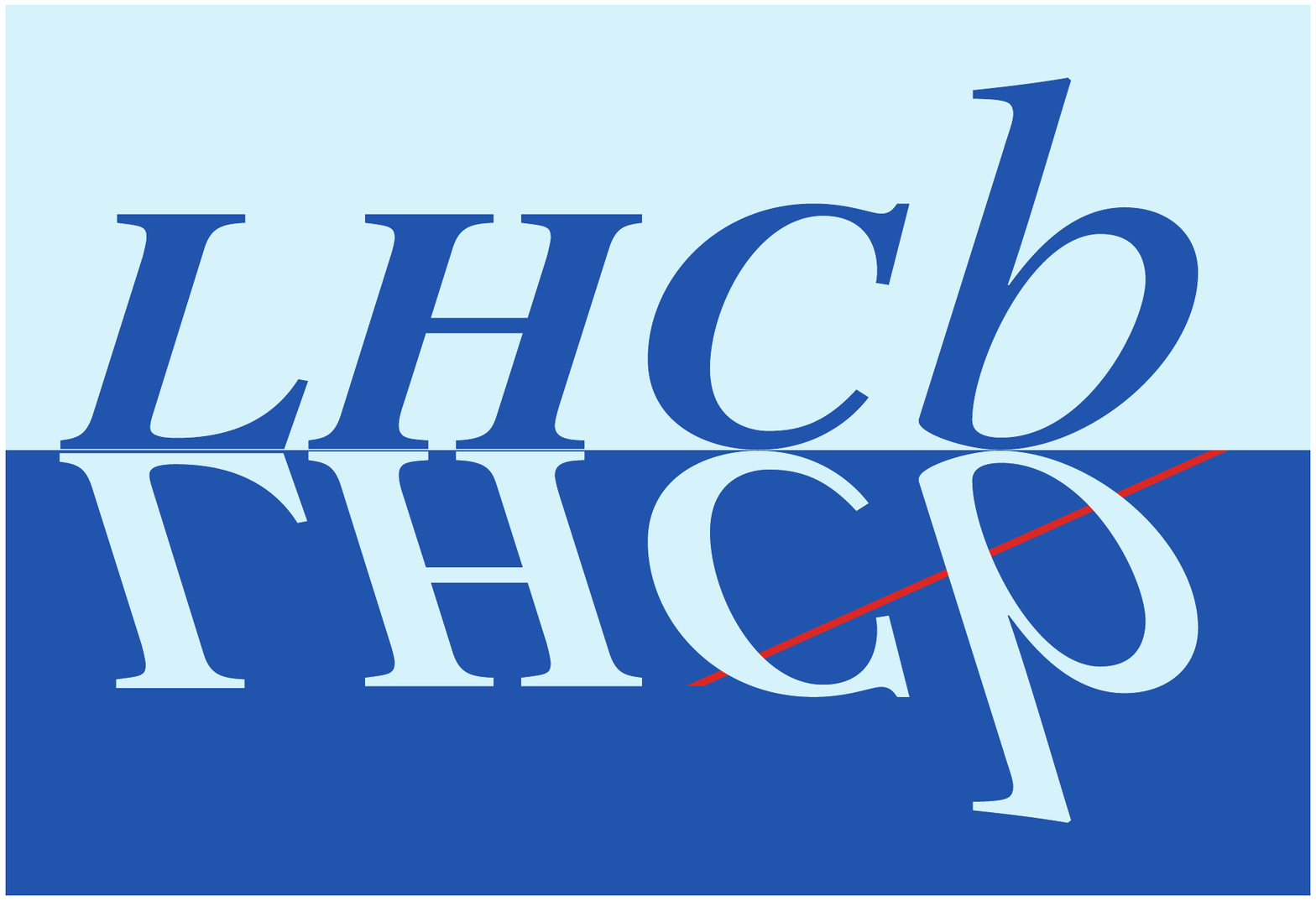}} & &}%
{\vspace*{-1.2cm}\mbox{\!\!\!\includegraphics[width=.12\textwidth]{lhcb-logo.eps}} & &}%
\\
 & & LHCb-DP-2018-001 \\  
 & & \today \\ 
 & & \\
\end{tabular*}

\vspace*{1.0cm}

{\normalfont\bfseries\boldmath\huge
\begin{center}
  Selection and processing of calibration samples to 
  measure the particle identification performance 
  of the LHCb experiment in Run 2
  \end{center}
}

\vspace*{0.5cm}

\def\markLAL{$^{1}$}
\def\markLPNHE{$^2$}
\def\markINFNBO{$^3$}
\def\markINFNCA{$^4$}
\def\markINFNFI{$^5$}
\def\markLNF{$^6$}
\def\markICCUB{$^{7}$}
\def\markCERN{$^8$}
\def\markNIKHEF{$^9$}
\def\markCAMBRIDGE{$^{10}$}
\def\markWAR{$^{11}$}
\def\markIMPE{$^{12}$}
\def\markOXFORD{$^{13}$}

\def\and{$^{,}$}

\begin{center}
R.~Aaij\markCERN, 
L.~Anderlini\markINFNFI, 
S.~Benson\markNIKHEF, 
M.~Cattaneo\markCERN, 
P.~Charpentier\markCERN,
M.~Clemencic\markCERN, 
A.~Falabella\markINFNBO,
F.~Ferrari,\markINFNBO,
M.~Fontana\markINFNCA\and\markCERN, 
V.~V.~Gligorov\markLPNHE, 
D.~Hill\markOXFORD,
T.~Humair\markIMPE,
C.~R.~Jones\markCAMBRIDGE, 
O.~Lupton\markCERN, 
S.~Malde\markOXFORD, 
C.~Marin~Benito\markICCUB,
R.~Matev\markCERN,
A.~Pearce\markCERN,
A.~Poluektov\markWAR,
B.~Sciascia\markLNF, 
F.~Stagni\markCERN, 
R.~Vazquez~Gomez\markCERN,
Y.-X.~Zhang\markLAL

\begin{footnotesize}
  \textit{
  \\\markLAL{Laboratoire de l'Accelerateur Lineaire, Paris, France}
  \\\markLPNHE{LPNHE, Universit\'{e} Pierre et Marie Curie, Universit\'{e} Paris Diderot, CNRS/IN2P3, Paris, France}
  \\\markINFNBO{Istituto Nazionale di Fisica Nucleare, Sezione di Bologna, Italy}
  \\\markINFNCA{Istituto Nazionale di Fisica Nucleare, Sezione di Cagliari, Italy}
  \\\markINFNFI{Istituto Nazionale di Fisica Nucleare, Sezione di Firenze, Italy}
  \\\markLNF{Istituto Nazionale di Fisica Nucleare, Laboratori Nazionali di Frascati, Italy}
  \\\markICCUB{ICCUB, Universitat de Barcelona, Barcelona, Spain}
  \\\markCERN{European Organization for Nuclear Research (CERN), Meyrin, Switzerland}
  \\\markNIKHEF{Nikhef, Amsterdam, Netherlands}
  \\\markCAMBRIDGE{University of Cambridge, Cambridge, United Kingdom}
  \\\markWAR{Department of Physics, University of Warwick, Coventry, United Kingdom}
  \\\markIMPE{Imperial College London, London, United Kingdom}
  \\\markOXFORD{University of Oxford, Oxford, United Kingdom}
  }
\end{footnotesize}

\vspace{\fill}

\newpage
\vspace*{1.0cm}
\begin{abstract}
  \noindent
  Since 2015, with the restart of the LHC for its second run of data taking,
  the LHCb experiment has been empowered with a dedicated computing model
  to select and analyse calibration samples to measure the performance
  of the particle identification (PID) detectors and algorithms.
  The novel technique was developed within the framework of the innovative 
  trigger model of the LHCb experiment, which relies on online event reconstruction for most of the 
  datasets, reserving offline reconstruction to special physics cases. The strategy to select and process
  the calibration samples, which includes a dedicated data-processing scheme combining online
  and offline reconstruction, is discussed. The use of the calibration samples to measure the detector PID performance, 
  and the efficiency of PID requirements across a large range of decay channels, is described. Applications of the calibration samples 
  in data-quality monitoring and validation procedures are also detailed.
\end{abstract}
\end{center}
\vspace*{0.2cm}

\begin{center}
  Submitted to EPJ TI
\end{center}

\vspace{\fill}

{\footnotesize 
\centerline{\copyright~CERN on behalf of the \lhcb collaboration, licence \href{http://creativecommons.org/licenses/by/4.0/}{CC-BY-4.0}.}}
\vspace*{2mm}

\end{titlepage}


\newpage
\setcounter{page}{2}
\mbox{~}
%
%
%
%

\cleardoublepage


\renewcommand{\thefootnote}{\arabic{footnote}}
\setcounter{footnote}{0}



\pagestyle{plain} 
\setcounter{page}{1}
\pagenumbering{arabic}


%


\section{Introduction}
\label{sec:Introduction}

LHCb is a dedicated heavy flavour physics experiment at the LHC. Its main goal
is to search for indirect evidence of new physics in CP-violating processes and rare decays
of beauty and charm hadrons.
Among other performance metrics, like excellent vertex resolution and 
good momentum and invariant-mass resolution, charged particle identification (PID)
distinguishing electrons, muons, pions, kaons and protons traversing the detector
is essential in the LHCb physics programme.
The required performances range from the per mille misidentification probability
of hadrons as muons in the study of the rare $B_{(d,s)}\to \mu^+\mu^-$ decays
\cite{LHCb-PAPER-2011-004%
,LHCb-PAPER-2011-025%
,LHCb-PAPER-2012-007%
,LHCb-PAPER-2012-043%
,LHCb-PAPER-2013-046%
,LHCb-PAPER-2014-049%
,LHCb-PAPER-2017-001%
}, to the sub percent precision, over a wide kinematic range accurate, on the detector induced asymmetries for the ambitious 
programme of CP asymmetry measurements 
\cite{LHCb-PAPER-2013-020%
,LHCb-PAPER-2014-017%
,LHCb-PAPER-2014-041%
,LHCb-PAPER-2017-021%
}.

PID information is extensively used both in the trigger selection and in offline
data analysis. This required the development of a dedicated computing model and a
strategy to select suitable calibration samples, in order to measure the PID performance and 
assess systematic effects. 

In Section \ref{sec:detector}, an overview of the LHCb detector is given, together with a summary of the PID calibration samples required in order to accomplish the physics goals of LHCb with Run 2 data (2015$-$2018). 
The article then focuses on the strategy to select and process 
PID calibration samples, including a description of the 
multivariate classifiers used to combine the response of calorimeters, 
RICH and muon system (Section \ref{sec:globalpid}); 
the procedure to measure the PID performance
using dedicated calibration samples, together with
the techniques to determine the selection efficiency
on hundreds of different decay channels, relying on a small number of calibration samples (Section \ref{sec:calib_samples});
the dedicated data-processing scheme 
combining online and offline reconstruction (Section \ref{sec:TurCal}); 
and the applications of the calibration samples to 
data-quality monitoring and validation (Section \ref{sec:validation}).
A brief summary and outlook are given in Section \ref{sec:conclusions}.
While this article discusses the calibration samples specifically for charged particle identification,
the general computing model and selection strategy is also being applied to other calibration samples in Run 2, such as those for tracking calibration and neutral pion and photon PID.

\section{Detector}
\label{sec:detector}

The \lhcb detector is a single-arm forward
spectrometer covering the \mbox{pseudorapidity} range $2<\eta <5$,
designed for the study of particlses containing \bquark or \cquark
quarks~\cite{Alves:2008zz,LHCb-DP-2014-002}. 
The detector includes a high-precision tracking system
consisting of a silicon-strip vertex detector surrounding the $pp$
interaction region~\cite{LHCb-DP-2014-001}, a large-area silicon-strip detector located
upstream of a dipole magnet with a bending power of about
$4{\mathrm{\,Tm}}$, and three stations of silicon-strip detectors and straw
drift tubes~\cite{LHCb-DP-2013-003} placed downstream of the magnet.
The tracking system provides a measurement of momentum, \ptot, of charged particles with
a relative uncertainty that varies from 0.5\% at low momentum to 1.0\% at 200\gevc.
The minimum distance of a track to a primary vertex, the impact parameter, 
is measured with a resolution of $(15+29/\pt)\mum$,
where \pt is the component of the momentum transverse to the beam, in\,\gevc.
Photons, electrons and hadrons are identified by a calorimeter system consisting of
scintillating-pad and preshower detectors, an electromagnetic
calorimeter and a hadronic calorimeter. 
Different types of charged hadrons are distinguished using information
from two ring-imaging Cherenkov (RICH) detectors~\cite{LHCb-DP-2012-003}. 
Muons are identified by a
system composed of alternating layers of iron and multiwire
proportional chambers~\cite{LHCb-DP-2012-002}.

The online event selection is performed by a trigger~\cite{LHCb-DP-2012-004},
which consists of a hardware stage, based on information from the calorimeter and muon
systems, followed by a software stage, which applies a full event
reconstruction.
Since 2015, in between the hardware and software
stages, a real-time procedure aiming at the alignment and calibration 
of the detector is performed \cite{LHCb-PROC-2015-011}, making use of a disk buffer~\cite{1742-6596-664-8-082011}.
Updated calibration parameters are made
available for the online reconstruction, used in the trigger selection.
Online calibration is of such high quality that it is also used for offline reconstruction, ensuring consistency between online and offline.

The responses of the calorimeter, \rich, and muon systems, or their combinations,
associated to each track in the reconstruction process are
named for brevity PID variables. 
They can be used in selections to increase the signal purity of a sample, reducing the processing time devoted  to the reconstruction of background events, often characterized by high multiplicity, and helping in fitting into the  data storage constraints.
Moreover they allow selections to avoid an explicit bias on quantities of physical
interest, such as decay time~\cite{LHCb-PAPER-2013-063,LHCb-PAPER-2013-065}.

The many contexts in which particle identification is exploited within the 
experiment and the difficulties in obtaining a perfect simulation for the PID 
detectors,
motivate the development of 
techniques for measuring the PID performance in
suitable PID calibration samples.
These samples are datasets collected by LHCb where decay candidates
have a kinematic structure that allows unambiguous identification of one of the 
daughters, without the use of any PID
information from the calorimeter, \rich, or muon systems, so that they are unbiased from the particle identification point of view.  
Today, most LHCb physics analyses rely on calibration samples for the determination 
of PID efficiencies.
In addition, these samples can be used to monitor time variations in performance, 
and to test new reconstruction algorithms.

The majority of physics analyses using data collected with the \lhcb
experiment rely on the physics quantities
as reconstructed in the online trigger reconstruction.
Still, physics analyses with special needs in terms of event reconstruction, 
searching for example for interactions of light nuclei or 
particles beyond the Standard Model with the detector \cite{LHCb-PAPER-2015-002},
are able to reprocess offline the collected calibration datasets with 
dedicated reconstruction algorithms.

In order to enable the measurement of selection efficiencies that
combine trigger requirements on the online-computed PID variables and offline requirements
of PID variables obtained through dedicated reconstruction algorithms, an innovative dedicated
data-processing strategy has been designed. 
Calibration data are obtained through a real-time selection based on the online 
reconstruction without any requirement on PID variables.
Each event belonging to the calibration samples is fully reconstructed independently 
both online and offline. The resulting reconstructed particles
are then matched, allowing a measurement of the efficiency of requirements that combine 
the two reconstruction types as described in Section \ref{sec:calib_samples}.


\section{Global particle identification}
\label{sec:globalpid}
The reconstruction algorithms of each of the PID
detectors of the LHCb experiment are very different, but each
of them allows the computation of a likelihood ratio between 
particle hypotheses for each reconstructed track.
The reconstruction algorithm of the RICH detectors provides
the likelihood of the electron, muon, kaon, proton and deuteron 
hypotheses relative to the pion hypothesis.
The calorimeter system provides the likelihood of 
electrons relative to the pion hypothesis.
Finally, the muon system provides the likelihoods 
of the muon and non-muon hypotheses.  
The likelihood ratios of the three detector systems are combined 
into \emph{Combined Differential Log-Likelihoods} (CombDLL)~\cite{LHCb-DP-2014-002}, which 
are used to define the selection criteria for the data analyses.
Selection strategies based on CombDLL and \isMuon~\cite{LHCB-DP-2013-001}, a binary
variable loosely identifying muons, are widely employed already
at the trigger level~\cite{Archilli:2013npa}.

Following recent developments in machine learning, more
advanced classifiers have also been designed to combine the likelihoods
ratios defined above with the informations from 
the tracking system, including the kinematic variables of the particle,
and additional information from the PID detectors not entering the 
likelihood computation (\emph{e.g.} the number of hits in the muon
system shared among reconstructed tracks).
The classifier with the widest application in this category, 
named \emph{ANNPID}, was developed using 
\emph{Forward Feeding Artificial Neural Networks}~\cite{Hocker:2007ht}, structured as a 
\emph{Multi-Layer Perceptron} (MLP) with a single hidden layer composed
of roughly 20\% more nodes than the input layer activated, through
a sigmoid function.
The training is performed minimising the likelihood of the 
Bernoulli Cross-Entropy with 
Stochastic Gradient Decent as implemented in the TMVA package.
The training sample is obtained from abundant simulated decays of
heavy hadrons 
that emulate the kinematic distributions of signal samples 
studied in several analyses.
Depending on the arrangement of the input samples, on the quality of 
the simulation, and on the available number of simulated events, the response
of the ANNPID algorithm can vary. As a consequence, 
the response of the ANNPID algorithms is provided in several tunings, some for 
general purpose, and others specialised for a particular analysis or kinematic range.
The variables combined using the ANNPID classifiers are listed in Table
\ref{tab:globalPID:ProbNNIN}.

\begin{table}
  \caption{\label{tab:globalPID:ProbNNIN}
    Input variables of the ANNPID classifiers for the various 
    subsystems of the LHCb detector.
    }
  \centering
  \begin{small}
  \begin{tabular}{ll}
    \hline
    \multicolumn{2}{l}{\emph{\textbf{Tracking}}}\\
     $\quad$ & Total momentum\\
             & Transverse momentum\\
             & Quality of the track fit\\
             & Number of clusters associated to the track \\
             & ANN response trained to reject ghost tracks \cite{GhostProb}\\
             & Quality of the fit matching track segments upstream and downstream of the magnet\\
    \hline
    \multicolumn{2}{l}{\emph{\textbf{RICH detectors}}}\\ 
     $\quad$ & Geometrical acceptance of the three radiators, depending on the direction of the track\\
             & Kinematical acceptance due to Cherenkov threshold for muons and kaons\\
             & Likelihood of the {electron}, {muon}, {kaon}, and {proton} hypotheses relative to the {pion}\\
             & Likelihood ratio of the {below-threshold} and pion hypotheses \\
    \hline
    \multicolumn{2}{l}{\emph{\textbf{Electromagnetic calorimeter}}}\\
     $\quad$ & Likelihood ratio of the {electron} and {hadron} hypotheses\\
             & Likelihood ratio of the {muon} and hadron hypotheses\\
             & Matching of the track with the clusters in the \emph{preshower} detector\\
             & Likelihood ratio of the electron and pion hypotheses, \\ & $\quad$after recovery of the Bremsstrahlung photons\\
    \hline
    \multicolumn{2}{l}{\emph{\textbf{Hadronic calorimeter}}}\\
     $\quad$ & Likelihood ratio of the {electron} and {hadron} hypotheses\\
             & Likelihood ratio of the {muon} and {hadron} hypotheses\\
    \hline
    \multicolumn{2}{l}{\textbf{\emph{Muon system}}}\\
             & Geometrical acceptance \\
             & Loose binary requirement already available in the hardware trigger\\
             & Likelihood of the {muon} hypothesis \\
             & Likelihood of the {non-muon} hypothesis \\
             & Number of clusters associated to at least another tracks\\
    \hline
  \end{tabular}
  \end{small}
\end{table}

All of the input variables for the ANNPID classifiers are made immediately available
to physics analyses, easing the development of new tunings and classification
algorithms dedicated to single analyses.
The many output variables of the detector reconstruction which are not used as input 
to ANNPID can be accessed or even regenerated, relying on the 
raw detector data stored on tape.

\section{Measuring PID performance}
\label{sec:calib_samples}
More than twenty exclusive 
trigger selections are designed to select pure samples of the five most common
charged particle species that interact with the LHCb detector: protons, kaons, pions, muons and electrons~\cite{Anderlini:samples}.
Generally, low-multiplicity decay modes with large branching fractions
are chosen in order to enhance the statistics and the purity and populate the tails in the 
distributions of the PID variables, which are of great relevance when computing misidentification
probabilities. Completely reconstructed final states composed of charged particles only are preferred, as they are selected with high purity at LHCb. An overview of the modes utilised is given in Table~\ref{tab:speciesregimes}.

The assumption underlying the usage of the calibration samples is that the distribution of the particle identification variables is independent of the selection strategy. Simply avoiding explicit requirements on the PID variables is not sufficient to ensure this.
In fact, the hardware trigger relies on information from the \calo and muon systems to reduce
the rate at which the full detector is read out to around $1\mhz$, while
a first layer of the software trigger, running before the full event reconstruction,
includes dedicated selection algorithms to identify high \pt muons and 
muon pairs. 

In order to avoid a pre-selection introducing biases in the PID variables, the selection strategy of the calibration samples imposes requirements on the algorithms selecting the event in the previous trigger layers. 
Either the trigger algorithms do not rely on PID information, or the PID selection in the trigger is applied to one of the particles not used to measure the performance.

\begin{table}[b]
  \caption[Overview of decay modes used to select calibration samples.]{Overview
           of decay modes that are used to select calibration samples. The high momentum samples are primarily selected, while the low momentum samples are included to maximise the kinematic coverage as much as possible.”
           }
  \begin{center}
    \begin{tabular}{lll}
      \hline
      \multicolumn{1}{c}{\textbf{Species}} &
      \multicolumn{1}{c}{\textbf{Low momentum}} &
      \multicolumn{1}{c}{\textbf{High momentum}}\\[3mm]
      \epm & \multicolumn{2}{c}{\decay{\Bp}{\jpsi\Kp} with \decay{\jpsi}{\epem} } \\[2mm]
      \mupm & \decay{\Bp}{\jpsi\Kp} with \decay{\jpsi}{\mumu} $\qquad$& \decay{\jpsi}{\mumu} \\[2mm]
      \pipm & \decay{\KS}{\pip\pim} & \decay{\Dstarp}{\Dz\pip} with  %
                                      \decay{\Dz}{\Km\pip} \\[2mm]
      \Kpm & \decay{\Dsp}{\Pphi\pip} with  \decay{\Pphi}{\Kp\Km} & \decay{\Dstarp}{\Dz\pip} with %
                                        \decay{\Dz}{\Km\pip} \\[2mm]
      \proton, \antiproton & \decay{\Lzz}{\prp\pim} & \decay{\Lzz}{\prp\pim} ;  %
                                      \decay{\Lc}{\prp\Km\pip} \\ [2mm]
      \hline
    \end{tabular}
  \end{center}
  \label{tab:speciesregimes}
\end{table}

Several of the selection strategies are implemented according to the so-called 
\emph{tag-and-probe} model. 
Taking the \decay{\jpsi}{\mumu} decay as an example, 
the tag-and-probe selection strategy relies on a list of
well-identified tag muons of a certain charge and a list of probe tracks 
with opposite charge, selected avoiding any PID requirement.
These are combined to form muon pairs with invariant-mass consistent with the \jpsi mass, 
and are then filtered further on the basis of the quality of the fit 
of the decay vertex, to form the final sample.
To extend the \pt range of the muons in the calibration samples to lower values, 
where the background from low momentum pions is difficult to reduce, the \jpsi
candidates can be combined with
charged kaons to form $\Bu\to\jpsi\Kp$ candidates\footnote{%
  Charged-conjugated candidates are implicitly considered here and throughout the paper.
}, adding further 
kinematic constraints related to the $B$ decay to the final filtering.

Proton calibration samples are obtained from two different decay modes:
\decay{\Lzz}{\proton\pim} and \decay{\Lc}{\proton\Km\pip}. 
Since the visible \Lzz production cross section in \lhcb is very high,
the yield collected at the trigger level exceeds the needs in terms of 
statistical precision on the particle identification. This would pose 
severe challenges for data storage. Therefore, a large fraction of these signal candidates 
is discarded by running 
the selection only on a randomly selected fraction of the events.
In order to improve the kinematic coverage of the sample, 
the fraction of discarded events is defined differently in four bins of the proton
transverse momentum (\pt),
resulting in a higher retention rate
in the less-populated high-\pt region. 
The sample of \Lc decays is included to extend the \pt coverage of the \Lzz samples. 

An abundant calibration sample for pions is provided by the 
decay $\KS \to \pip\pim$, but 
the spectrum of the probe particles is much softer than what is 
typical for hadrons produced in heavy hadron decays.
Charm hadron decays allow the kinematic range to be extended to higher
transverse momenta, but the lower purity of the samples, due to 
the smaller production cross-section, requires additional care
in the selection and background subtraction strategies.The decay
\decay{\Dstarp}{\Dz\pip} with \decay{\Dz}{\Km\pip} represents
the primary source of \pipm and \Kpm calibration samples.
The soft pion produced in the strong decay of the \Dstarp hadron
allows to tag the flavour of the \Dz and therefore to
distinguish the kaon and the pion produced in its decay without 
PID requirements on either of the two probe particles.
Applying a requirement on the energy release in the $D^*$ decay, which is expected to be small, enables the rejection 
of combinatorial background due to the erroneous combination
of \Dz hadrons and pions produced in unrelated processes.
Finally, the \decay{\Dsp}{\Pphi\pip} decay with \decay{\Pphi}{\Kp\Km} 
is a further source of kaons. This sample allows the kinematic range for kaons to be extended to lower momenta, as the $\Pphi$ constraint enables the kinematic requirements on the kaons to be loosened while retaining the purity.

\begin{figure}
  \begin{center}
    \includegraphics[width=0.32\textwidth]{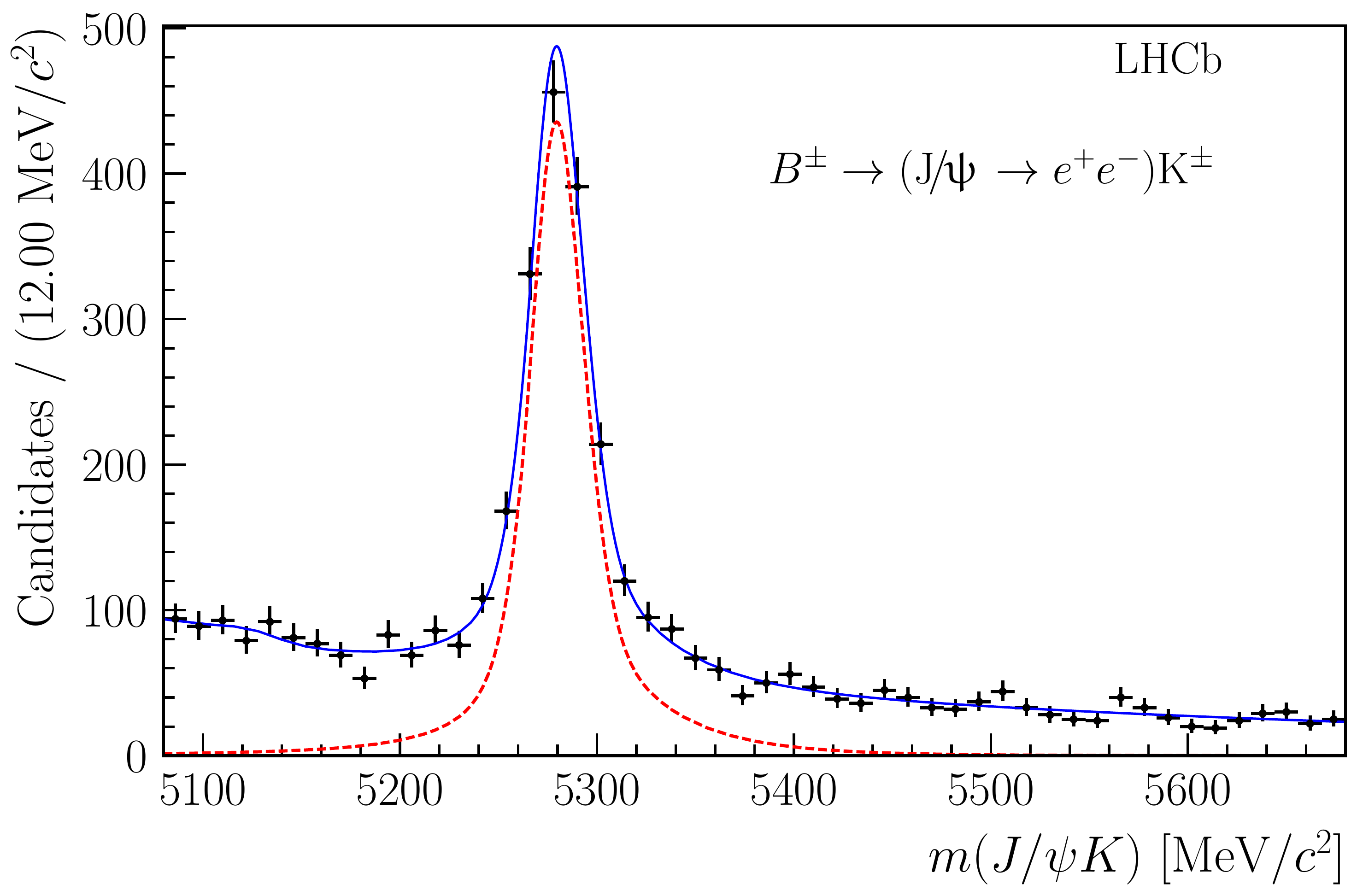} 
    \includegraphics[width=0.32\textwidth]{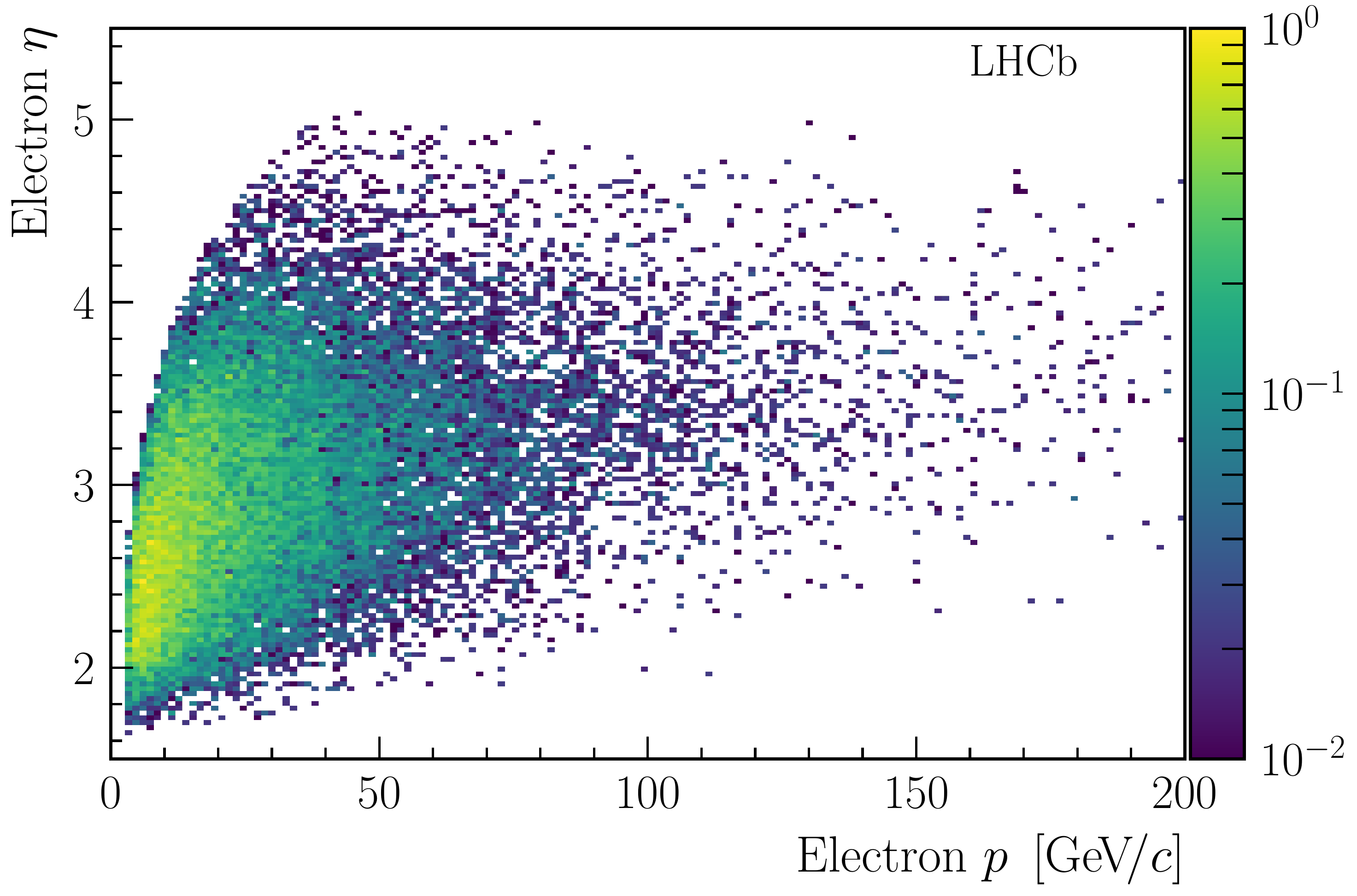}\\
    \includegraphics[width=0.32\textwidth]{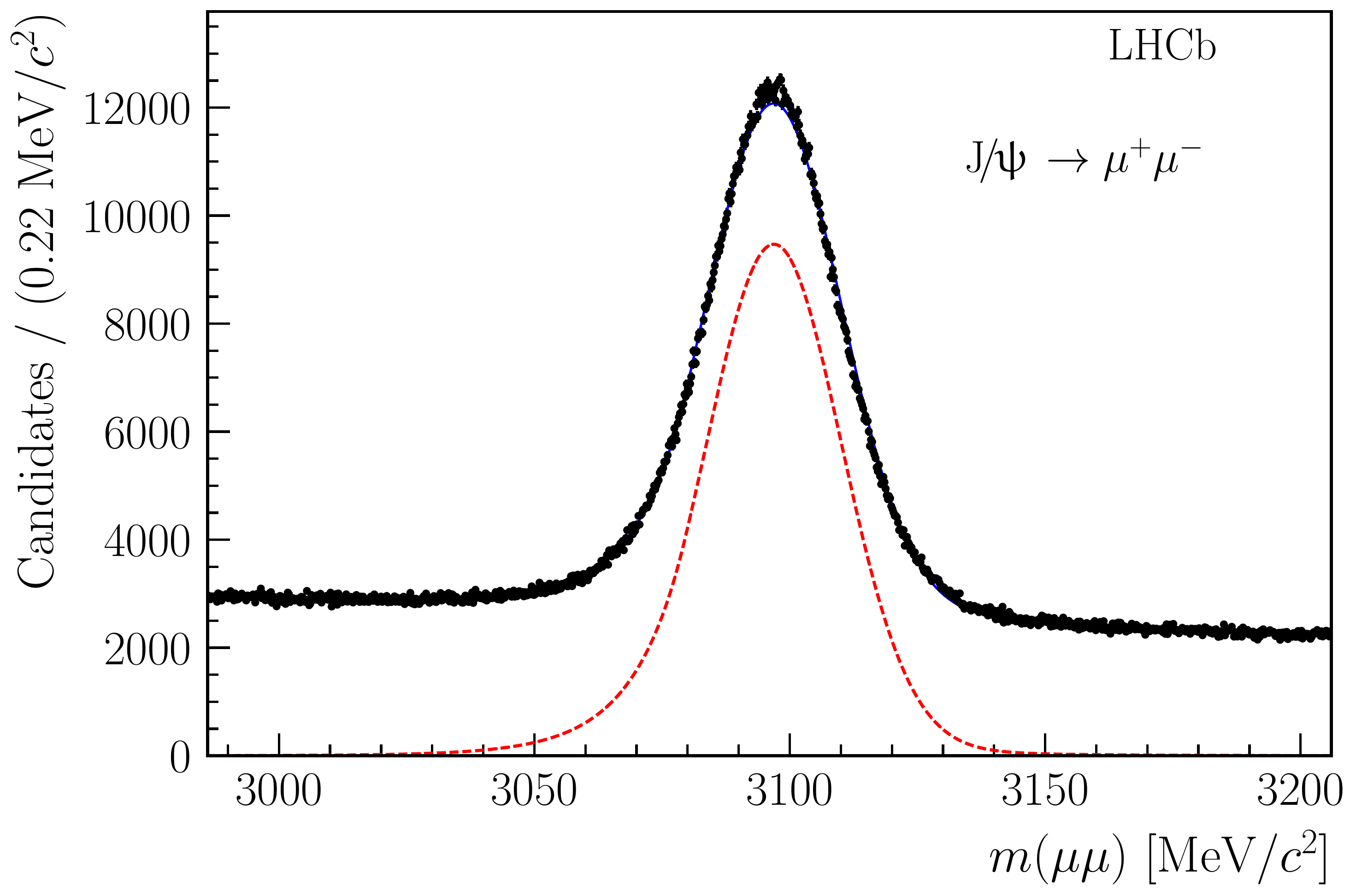} 
    \includegraphics[width=0.32\textwidth]{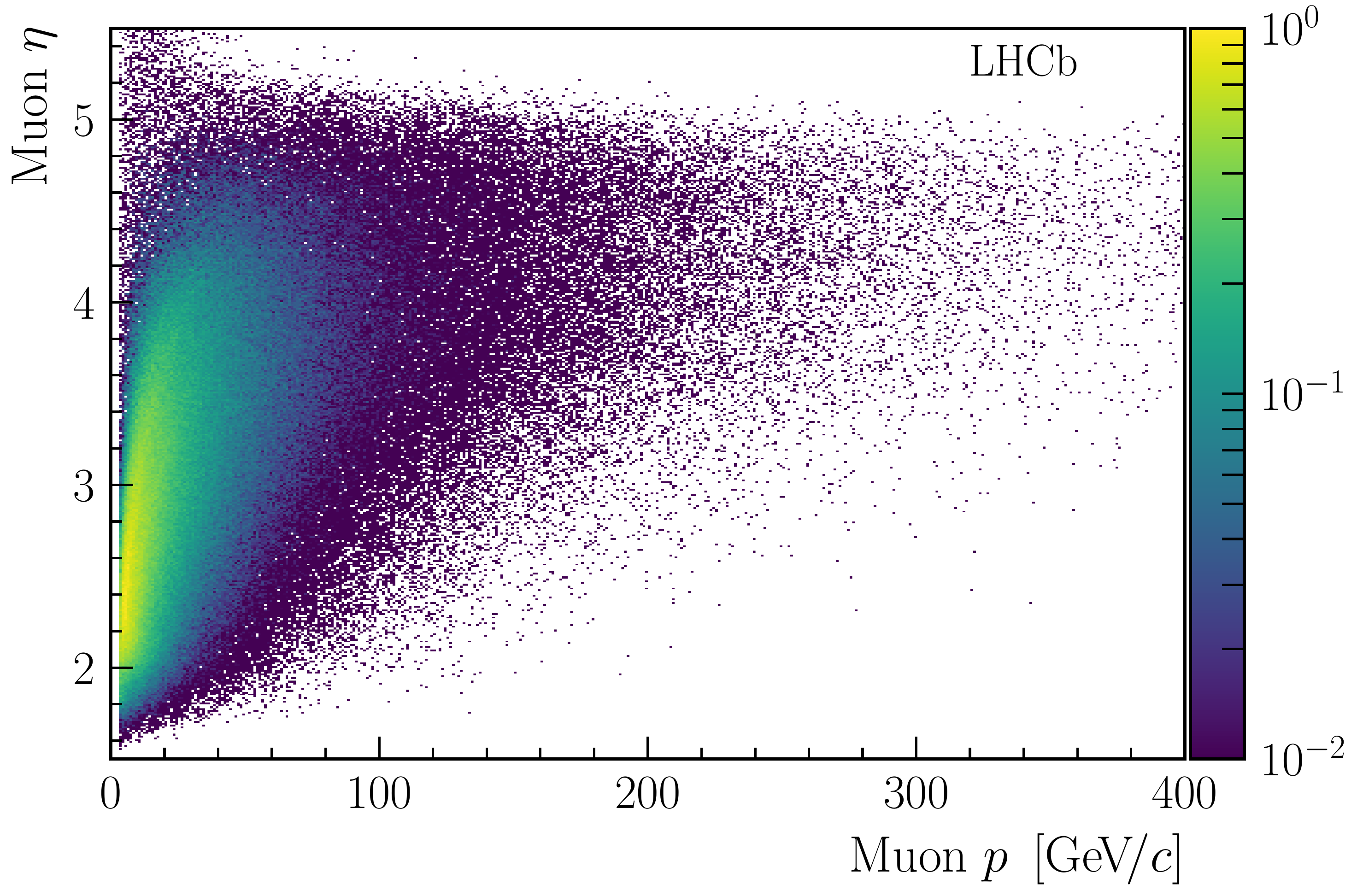}\\
    \includegraphics[width=0.32\textwidth]{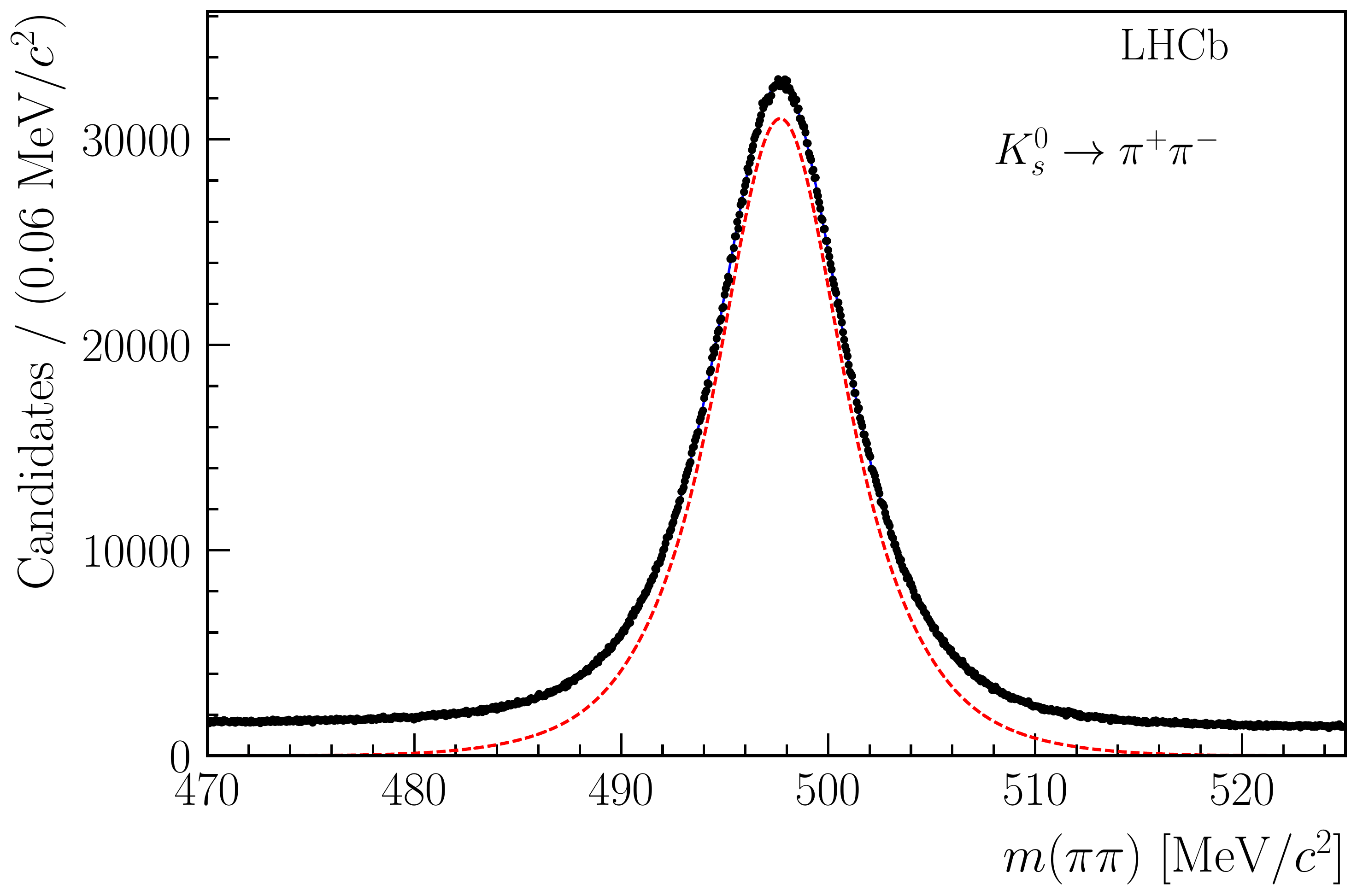} 
    \includegraphics[width=0.32\textwidth]{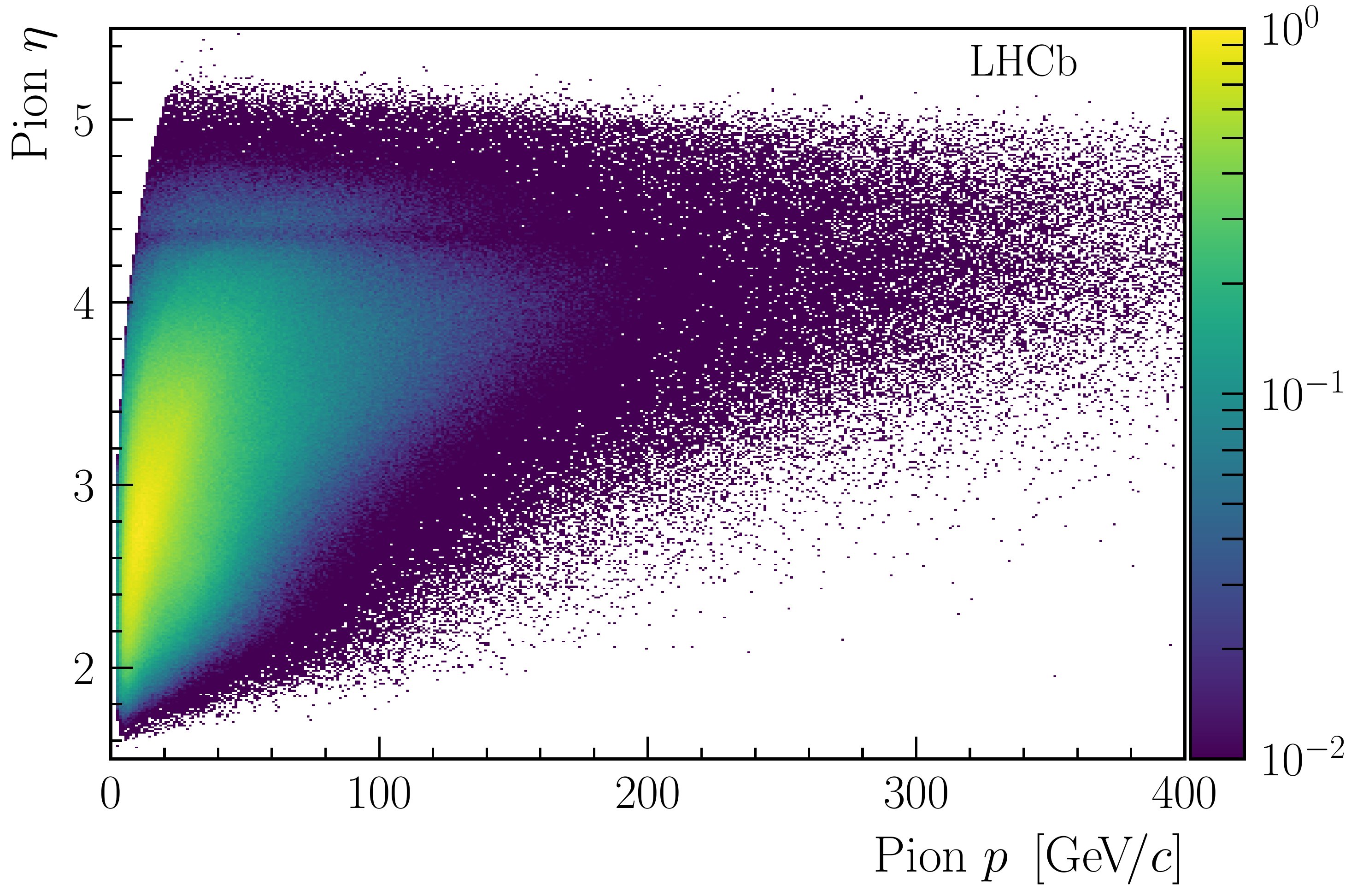}\\
    \includegraphics[width=0.32\textwidth]{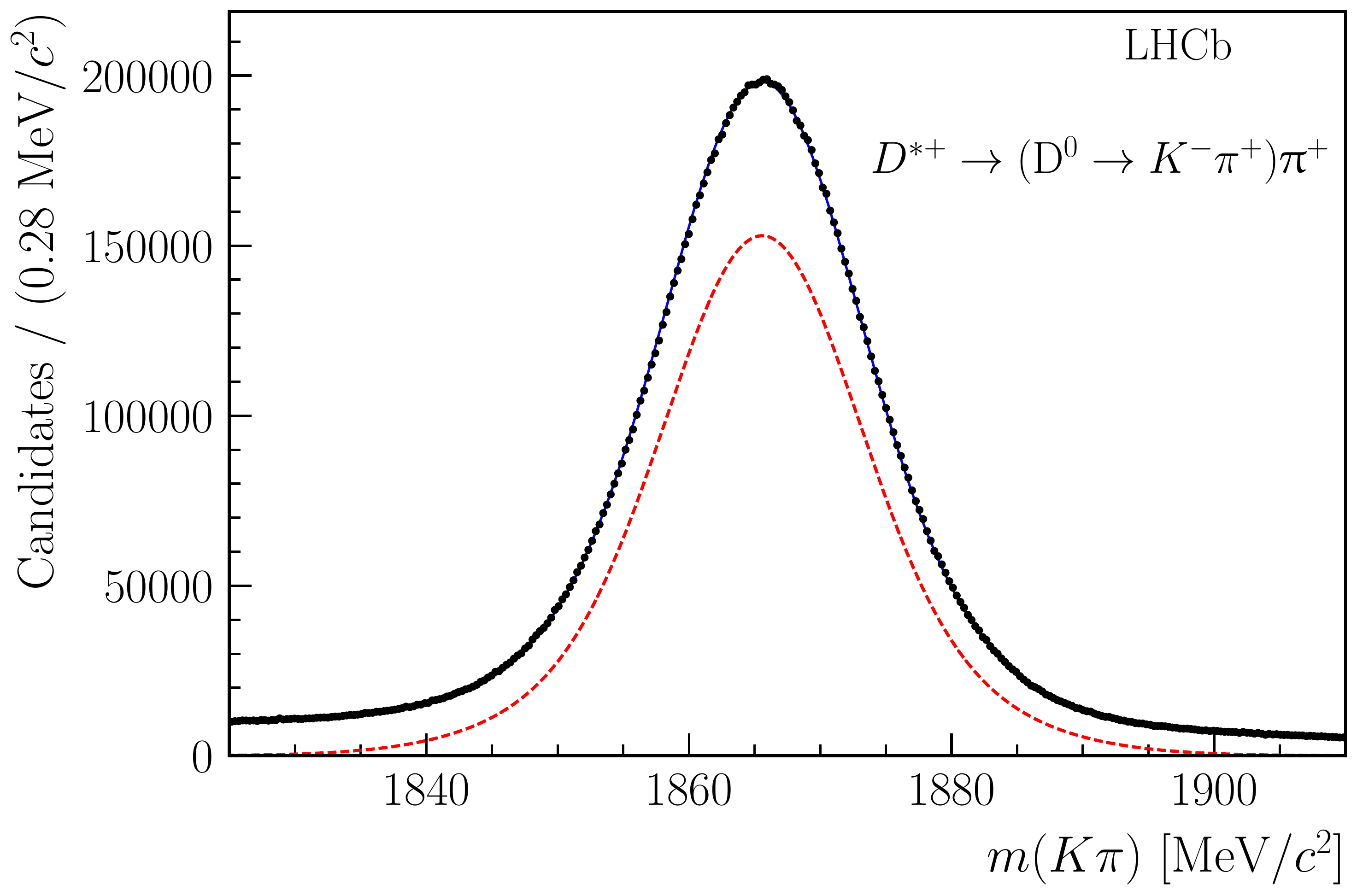} 
    \includegraphics[width=0.32\textwidth]{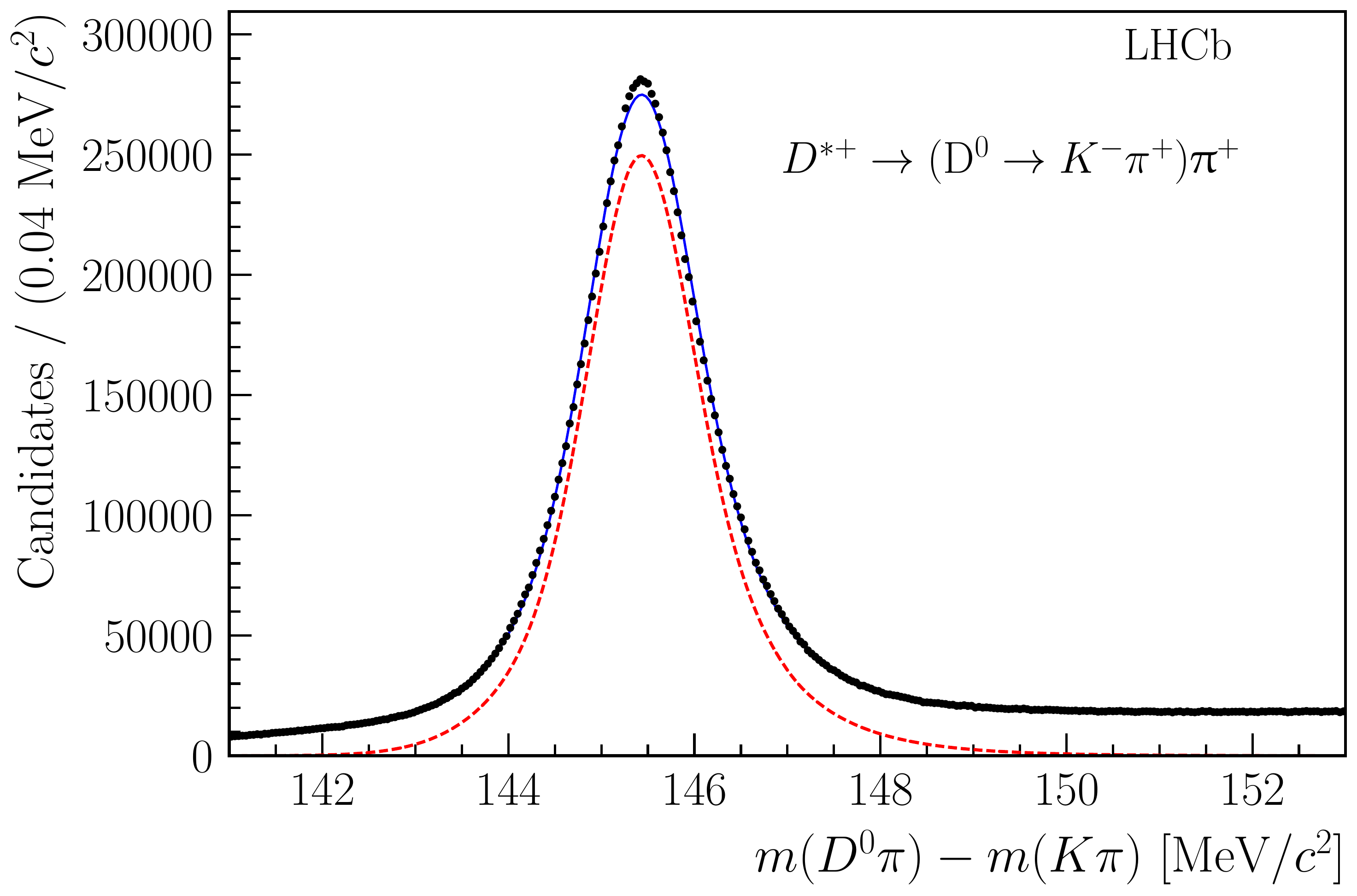} 
    \includegraphics[width=0.32\textwidth]{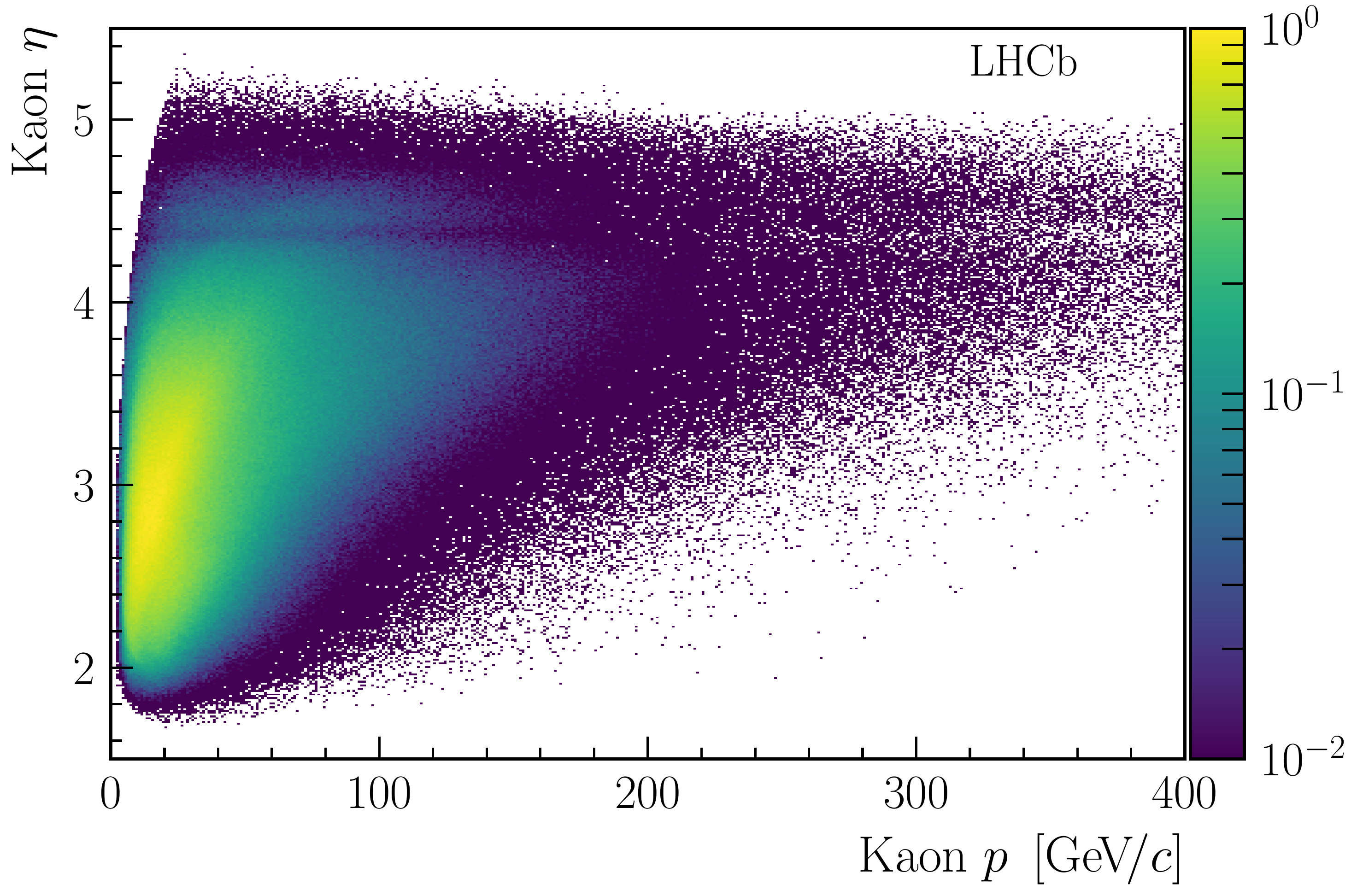}\\
    \includegraphics[width=0.32\textwidth]{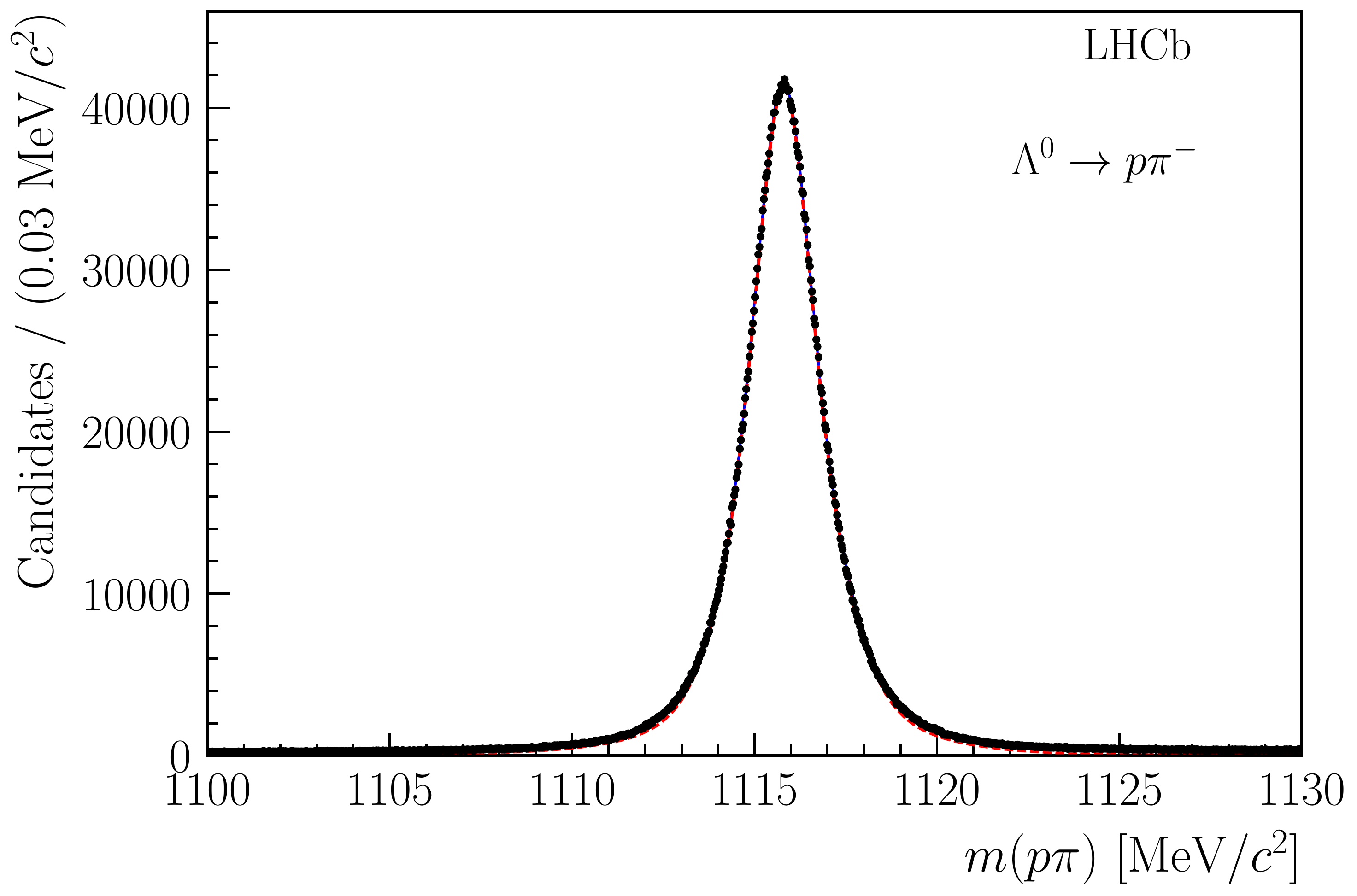} 
    \includegraphics[width=0.32\textwidth]{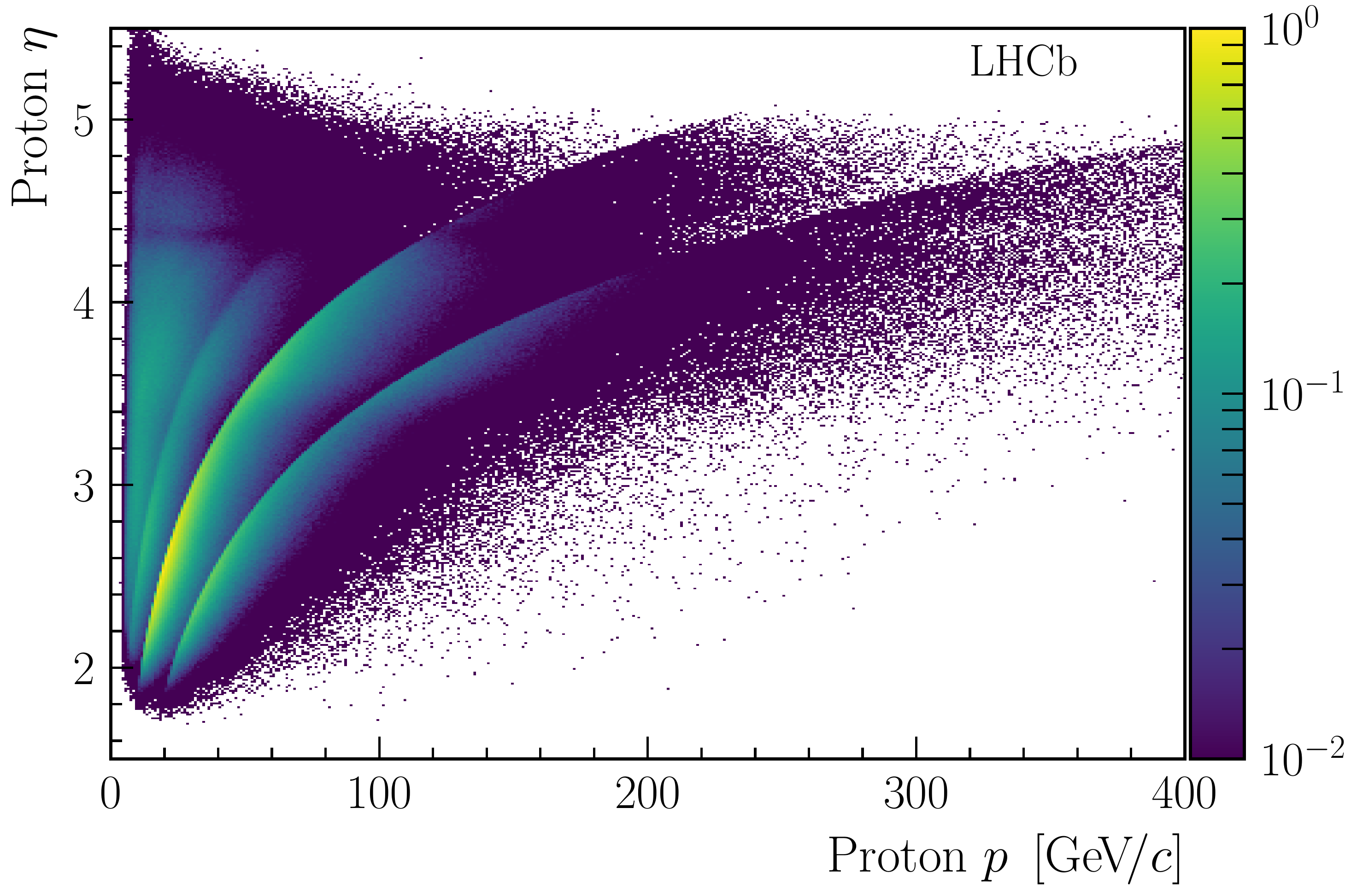}\\
  \end{center}
  \vskip-5mm
  \caption{\label{fig:all_particle_coverage}
    On the left, 
    mass distributions of the decaying particles with the results of the fit superimposed; signal contributions are shown
    by the red dashed curves, and the total fit functions including background contributions are shown by the blue solid curves. 
    On the right, the background-subtracted distributions of the calibration samples for
    electrons, muons, pions, kaons and protons
    as a function of the 
    track pseudorapidity, $\eta$, and momentum \ptot are shown.
    The colour scale units are arbitrary.
  }
\end{figure}

The residual background that cannot be rejected with an efficient
selection strategy is statistically subtracted using the 
\sPlot technique~\cite{Pivk:2004ty}. 
A fit to the invariant-mass of the decaying particle is performed for each calibration sample, defining a signal component for which the sample of probe tracks is known to be pure, and one or more background components of different nature.  
In several cases, two-dimensional fits are performed to account for additional background sources. The variables used in the two-dimensional fits are: the $\Dz$ mass and $\Dstarp - \Dz$ mass difference for $\Dstarp \to \Dz \pip$; the $\Bp$ and $J/\psi$ masses for $\Bp \to (J/\psi \to \mup \mu^-) \Kp$;  the $\phi$ and $\Dsp$ masses for $\Dsp \to \phi \pip$.

The fit to each calibration sample
is repeated (at least) twice; the first iterations have a large
number of free parameters including the means and widths of the signal 
components and shape parameters of the background components, whereas 
the final iteration fixes all of the parameters apart from the normalisation of 
each component (such as signal, misidentified background and combinatorial 
background). The covariance matrix produced in the final fit is used to define a relation between the discriminant variables and a signal \sWeight to be assigned to the daughter candidate. Correlations between the chosen discriminating variables and the PID variables do not play a significant role.
Figure~\ref{fig:all_particle_coverage} illustrates the invariant-mass distributions for some of the calibration samples, as obtained from proton-proton collision data collected in 2015 corresponding to an integrated luminosity of 0.17\invfb. The corresponding kinematic distributions for the different species of probe particles are 
also shown.


The response of the PID detectors to a traversing particle depends on
the kinematics 
of the particle, the occupancy of the detectors (which may be different 
event-to-event and for different particle production mechanisms), 
and experimental conditions such 
as alignments, temperature, and gas pressure (which may modify the response of 
detectors across runs). 

One may assume that the response of a PID variable
is fully parameterised by some known set of variables, such as the track 
momentum \ptot (which is related to the Cherenkov angle in the RICH and to the energy deposited in the calorimeter ) and the track multiplicity, the latter being given by
the number of reconstructed 
tracks traversing the whole detector. 
By partitioning the sample with sufficient granularity
in these parameterising variables,
the PDF of the PID variable distribution does not vary 
significantly within each subset, such
that the efficiency of a selection requirement on that variable is constant within each subset.

In the trivial case of events that come from the calibration sample, there is no need to compute per-subset efficiencies, and the average efficiency is simply given by the fraction of background subtracted events passing the PID requirement. To compute the PID efficiency on a sample other than the calibration sample, 
denoted hereafter as the \emph{reference sample}, the parameterising variables in the 
calibration sample can be weighted to match those in the reference sample.
The PID efficiency can then be computed 
using the per-subset weights.
The weights are defined as the normalised ratio of reference to calibration 
tracks
\begin{equation}
  w_{i} = \frac{R_{i}}{C_{i}}\times\frac{C}{R}\; ,
  \label{eqn:calibration:weight}
\end{equation}
where $R_{i}$ ($C_{i}$) is the number of reference (calibration) tracks in the 
\ith subset, and $R$ ($C$) is the total number of reference (calibration) 
tracks in the sample.

After applying the PID cut to the weighted calibration sample, 
the average efficiency of the PID requirement on the weighted calibration 
sample is
\begin{equation}
  \avgeff = \frac{\sum_{i}\eff_{i}w_{i}C_{i}}{\sum_{i}w_{i}C_{i}}\; .
  \label{eqn:calibration:avgeff}
\end{equation}
where $w_{i}$ is the per-subset weight, $\epsilon_{i}$ is the per-subset 
efficiency and $C_{i}$ is the number of calibration tracks in the $i$-th subset. 

The computation of the PID efficiency can be thought of as the reweighting of the calibration sample to match the 
reference, or as the assignment of 
efficiencies to reference tracks based on the subset they belong to.
This can also be extended to reference samples where PID requirements have been 
imposed on multiple tracks, where the efficiency of an ensemble of 
cuts is required taking into account the kinematic correlation between tracks.

There are a number of ways in which the calibration samples can be used to determine PID efficiencies.
Three broad strategies have been commonly implemented by \lhcb in the past. The first uses a simulated reference sample to provide the kinematics of the signal tracks under consideration. This is an ideal approach to use when the kinematics of the signal tracks are known to be well modelled in the simulation. If the signal in data can be reliably separated from the other species in the sample, 
such that some background subtraction 
can be used to extract the signal kinematics, a second approach to creating the reference sample can be used.

Lastly, the PID response of MC signal samples can be corrected using
the PID calibration data samples. Two options are provided: 
\begin{itemize}
\item \textit{Resampling} of PID variables, where the PID response is completely replaced 
by the one statistically generated from calibration PDFs.
\item \textit{Transformation} of PID variables, where the PID variables from the 
simulation are transformed such that they are distributed as in data.
\end{itemize}

\noindent The PID correction is still considered as a function of track kinematics 
($\pt$ and $\eta$) and event multiplicity $N_{\rm evt}$ (such as the number of tracks in the event).
However, unlike in the first two strategies detailed above, the correction is performed using an unbinned approach, where 
the calibration PDFs in four dimensions, the PID variable, $\pt$, $\eta$, and a measure of $N_{\rm evt}$, are described 
by a kernel density estimation procedure using the Meerkat library~\cite{Poluektov:2014rxa}. 
The advantage of resampling and variable transformation is that the corrected PID response can be used as an input to a multivariate classifier.

However, a limitation of the PID resampling approach is that the PID 
variables for the same track are generated independently, and thus no 
correlations between them are reproduced. Therefore, only one PID variable 
per track can be used in the selection. Correlations between variables for 
different tracks are preserved via correlations with the kinematics
of tracks, assuming the PID response is fully parameterised by $\pt$, $\eta$, and $N_{\rm evt}$.

\begin{figure}
  \centering
  \includegraphics[width=.9\textwidth]{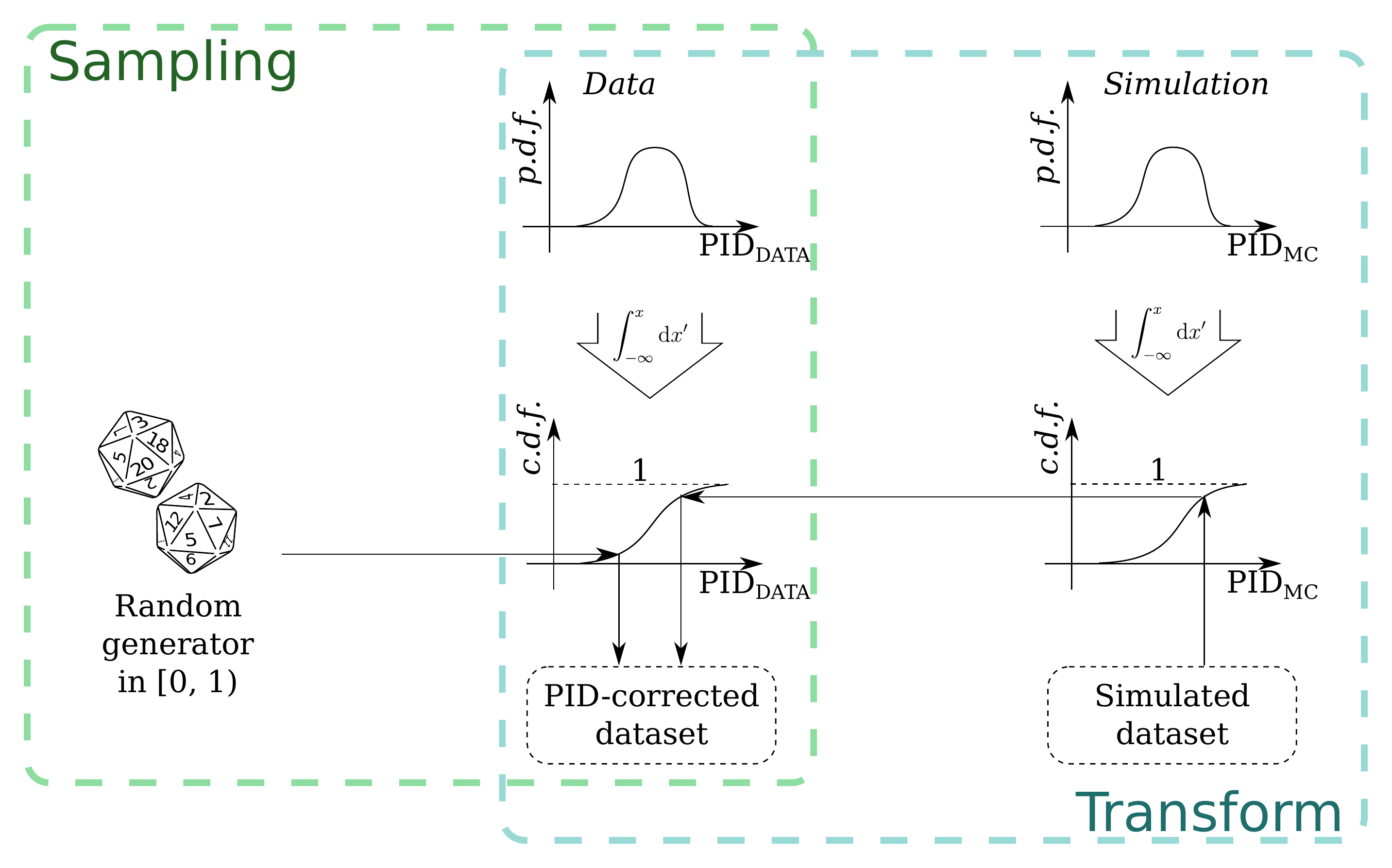}
  \caption{\label{fig:PIDGen}
    Schematic representation of the PID resampling and variable transformation
    techniques. 
  }
\end{figure}

The PID variable transformation approach aims to remove this limitation~\cite{Patrignani:2016xqp}. 
The corrected PID variable ${\rm PID}_{\rm corr}$ is obtained as 
\begin{equation}
  {\rm PID}_{\rm corr} = P^{-1}_{\rm exp}(P_{\rm MC}({\rm PID}_{\rm MC}|\pt, \eta, N_{\rm evt})|\pt, \eta, N_{\rm evt}), 
\end{equation}
where $P_{\rm MC}({\rm PID}_{\rm MC}|\pt, \eta, N_{\rm evt})$ is the cumulative 
distribution function of the simulated PID variable ${\rm PID}_{\rm MC}$, 
and $P^{-1}_{\rm exp}(x|\pt, \eta, N_{\rm evt})$ (where $0<x<1$) is the inverse cumulative 
distribution function for the PID variable from the calibration sample 
(\ie for fixed \pt, $\eta$ and $N_{\rm evt}$ it returns the PID variable that corresponds to 
a cumulative probability $x$). Both functions 
are obtained from the results of kernel density estimations of the simulation 
and calibration PID responses, respectively. 
The corrected PID variables obtained in this way follow the PDF of the calibration sample, 
but preserve strong correlations with the output of simulation. 
Through these correlations in simulation, the 
ones between PID variables for the same track are reproduced 
to first order. The drawback of this approach is that it also relies on the 
parametrisation of PID PDFs in simulation, which are extracted from samples
that are typically much smaller than the calibration data. 
Although one naively expects this method to perform better due to taking correlations into account, studies are ongoing to quantify the degree of agreement between the correlations found in simulation and data.
The PID resampling and variables transformation techniques are schematically 
represented in Figure \ref{fig:PIDGen}.

There are a number of sources of uncertainty that affect the measurement of 
PID efficiencies. The statistical uncertainty arises from finite statistics in 
the input samples used in the calibration procedure, namely the calibration 
and reference samples. Due to the large calibration sample sizes, this 
uncertainty is usually dominated by the size of the signal reference sample. 

Several sources of systematic uncertainty related to the procedure must also 
be accounted for, arising from differences between the reference and signal 
samples, the specific choice of binning used, and the \sWeight procedure used 
in the calibration sample production. The degree to which these uncertainties 
affect the PID efficiency precision is analysis dependent, and require 
specific studies to be carried out on a case-by-case basis.
Moreover the availability of primary and secondary calibration samples allows to study possible biases coming from single decay modes.

\section{Computing model for the calibration samples}
\label{sec:TurCal}

\begin{figure}
  \centering
  \includegraphics[height=0.5\textheight]{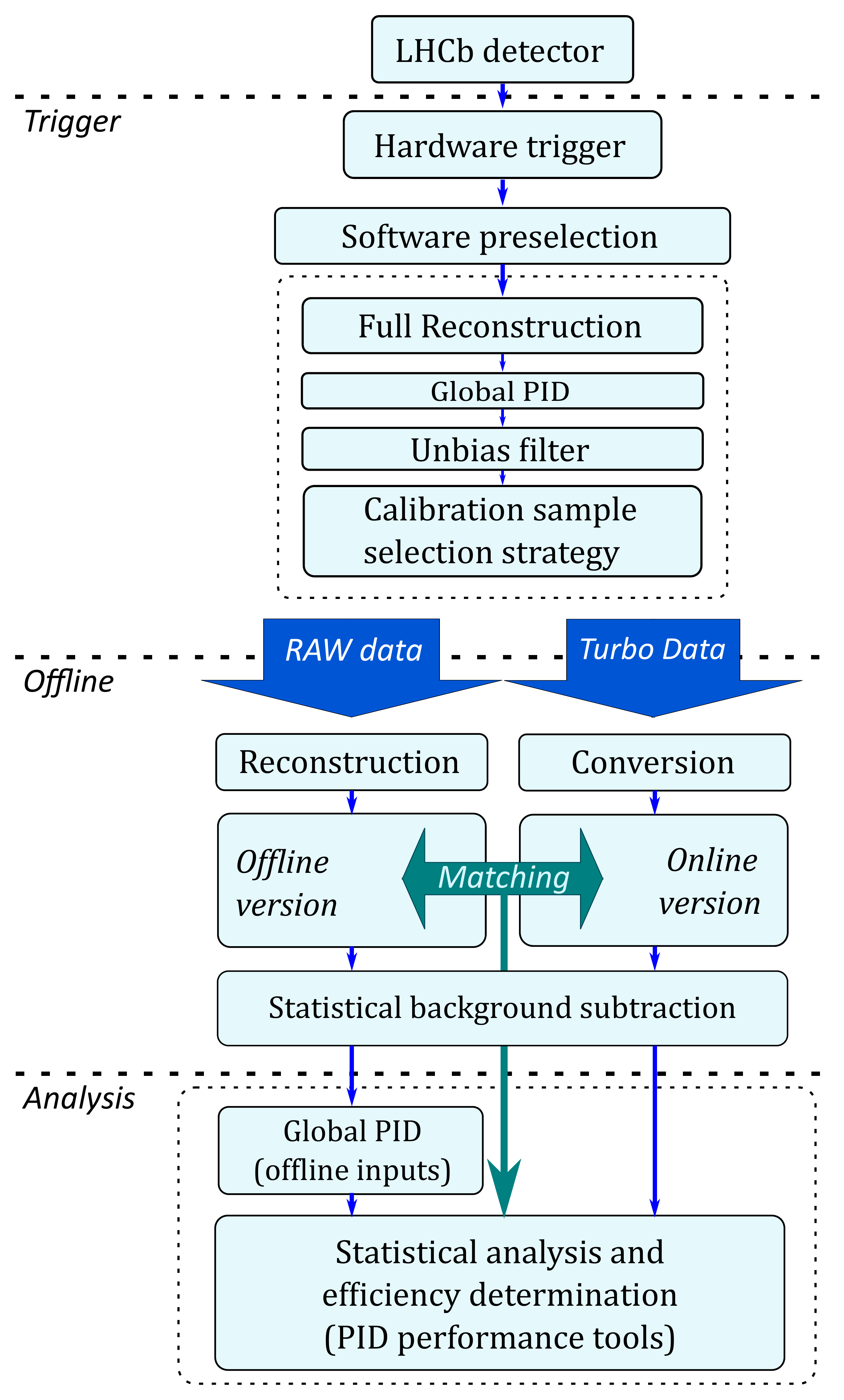}
  \caption{\label{fig:TurCal:computingModel}
    Schematic representation of the computing model for the 
    PID calibration samples.
  }
\end{figure}

In order to face the new challenges of the second run of the LHC, 
the LHCb trigger~\cite{Aaij:2012me,Albrecht:2013fba} has evolved into a 
heterogeneous configuration with 
different output data formats for different groups of trigger selections.
Figure \ref{fig:TurCal:computingModel} shows a schematic representation 
of the computing model that is described in the following.

Two alternative data formats for physics analyses, 
named \Turbo stream~\cite{Aaij:2016rxn} and \Full stream, have been developed.
Trigger selections writing to the \Turbo stream are intended for analyses 
of samples where only the information related to the candidates and associated reconstructed objects is needed.
Trigger selections that are part of the Turbo stream produce a decay candidate which is stored for offline analysis, along with a large number of detector-related variables, while the raw detector data is not kept~\cite{LHCb-DP-2016-001,CERN-LHCC-2014-016}.
When considering analyses based on the \Turbo stream, it is therefore evident that
the calibration samples must provide the PID information
as computed online in order to assess the efficiency of selection requirements applied 
either in the trigger selection, or offline on the PID variables retrieved from 
the trigger candidate. 

Trigger selections for events to be stored in
the \Full stream are intended for those measurements
and searches for which the Turbo approach is not applied. 
While the software trigger fully reconstructs candidates, 
those are not saved.
If the trigger decision is affirmative, the raw detector data
is saved together
with summary information on the trigger decision, including
the CombDLL and \texttt{isMuon} variables,
for each particle involved in the trigger decision.
The track and decay candidates are reproduced in a further offline 
reconstruction step that accesses the raw detector data.
Indeed, some physics data analyses present special needs in terms of 
particle identification algorithms, for example because they explore kinematic
regions at the boundaries of the kinematic acceptance, or because of
exceptional requirements in terms of the accuracy of the efficiency determination.
To respond to such special requirements, dedicated algorithms accessing the raw detector data
can be developed and included in the offline event reconstruction.
Hence, the events selected as part of the calibration samples must include
the raw data, allowing the performance of future algorithms 
to be measured on data.

An interesting case is presented when a trigger selection targeting the \Full
stream includes PID requirements that are then intended to be refined offline.
Potentially, the PID variables computed online can differ from those obtained 
from the full event reconstruction performed offline. While accidental differences 
in the online and offline algorithms are unlikely thanks to dedicated checks in the 
data quality validation procedure, the offline reconstruction is subject to improvements
that provide a slightly different value for the PID variables.
The determination of the efficiency of combined requirements on 
online and offline versions of the PID variables, or of different tunings of the
multivariate classifiers adopted in the trigger and in the statistical data analysis,
require the use of calibration samples combining the information from 
the online and offline reconstruction, allowing full offline reprocessing if needed.

A dedicated data format, named \TurboCalib, was developed to satisfy the requirements on PID
calibration samples described above~\cite{Anderlini:2199780}.
After the online full event reconstruction, events in which decay candidates 
useful for calibration are identified and selected in real-time
are stored including both the trigger candidates themselves and raw
detector data. The two output formats are processed independently for each event, to obtain 
both decay candidates propagated from the trigger and decay candidates 
reconstructed offline from the raw detector data.
The two reconstructions are fully independent, so that
the tracks identified in the two processes must be matched. This is done according to the 
fraction of shared clusters in the detector, or exploiting the TisTos algorithm 
described in Ref.~\cite{TisTos}, or with a combination of the two techniques. 

The offline versions of the PID variables can be easily replaced with 
other tunings of the multivariate classifiers, or through the output of
dedicated reconstruction sequences. As a result of the matching procedure, 
each reconstructed track is associated to two sets of PID variables, 
obtained through the online and offline versions of the reconstruction, respectively. 
The two sets are available to the analysts to measure the efficiency
of selection requirements that possibly combine the two versions.

As described in Section \ref{sec:calib_samples},
the measurement of the selection efficiencies from the selected 
calibration samples is enabled through the subtraction of 
the residual background by means 
of the \sPlot technique.
In order to overcome to the scalability challenges set by the increasing needs for 
precision in many LHCb measurements, resulting in huge calibration samples to control
the statistical uncertainty, the background subtraction is 
performed through a dedicated, distributed implementation of the \sPlot technique.
Finely binned histograms of the invariant-mass distributions of the trigger candidates are filled in parallel on thousands of computing nodes.
They are then merged and modeled through a maximum likelihood fit as
the combination of signal and background components. 
The relations between the discriminating variables and the \sWeights to be assigned to each candidate 
are sampled in fine grids and made available through a distributed file system to the computing 
nodes of the LHCb grid~\cite{DIRAC}, where jobs to assign the weights are run as a final
processing step in the calibration sample production workflow.
Such a distributed implementation of the \sPlot technique avoids 
the storage of the entire dataset on a single computing node, hence scaling 
better with the size of the calibration samples.

The real-time selection strategy, 
the double-processing scheme combining event-by-event the online and 
offline reconstructed variables, and the distributed approach to 
background subtraction constitute the main novelties in the data 
processing for the calibration samples, overcoming most scalability 
issues and making the limited cross-section and the available data 
storage resources the only limitations to the statistical precision in the
determination of PID selection efficiencies.

Finally, the \emph{Particle IDentification Calibration} 
(PIDCalib) package~\cite{Anderlini:2202412} is a user interface 
written in python aiming at a standardization of the techniques 
described in Section \ref{sec:calib_samples} to transfer the 
information on PID of the calibration samples to the reference
sample of interest for the many physics analyses. 
It includes several reweighing approaches, PID resampling and 
PID variable transformation. The set of variables identifying 
the kinematics of the tracks, the event multiplicity and the PID
response can be chosen case by case, while the access to calibration
samples and the implementation of the algorithms are maintained centrally.

\section{Data quality, monitoring and validation}
\label{sec:validation}
As discussed in Section \ref{sec:calib_samples}, calibration samples are 
abundant decays selected at the trigger level with high purity, representative
of all the families of long-lived particles interacting with the \lhcb 
detector, except deuterons.
Their immediate availability during the data taking and their high 
statistics are key ingredients for data-quality monitoring and validation.
Since the reconstruction involves different systems of the \lhcb detectors
depending on the nature of the particle, the various samples are used to 
monitor and validate different aspects of the reconstruction. For example,
the recovery of Bremsstrahlung photons to improve the momentum resolution
of the electrons can only be monitored and validated using an electron sample.
Similarly, the efficiency of muon identification can be better monitored and 
validated using a sample of tagged muons.

In order to add redundancy to the validation procedure, a small fraction
of the calibration samples are reconstructed with the offline procedure in real 
time, in order to trigger alarms in a timely manner in case of misalignments 
between the online and offline reconstruction, due for example to errors in the 
database handling the alignment and calibration constants.

Finally, several checks on the reconstructed quantities of the calibration samples 
have been included in the automated validation procedure performed during data taking, aiming 
at an early identification of deviations from standard running conditions, 
and to check possible temporal variations in performance due to unstable 
environmental conditions, or ageing of the detector \cite{DQM}.

Real-time monitoring on pure decay samples representative of the needs of a 
wide physics programme will be of critical importance during Run 3 of the LHC, 
when, after a major upgrade of the LHCb experiment, most datasets to perform 
physics data analyses will be selected in the trigger and stored as decay 
candidates, with no support for raw detector data \cite{CERN-LHCC-2014-016}. 
Since no further reprocessing of the reconstruction will be possible, any
loss in performance will unavoidably result in a loss of effectiveness for the resulting 
physics measurements.


\section{Conclusions}
\label{sec:conclusions}

The strategy to select and process the calibration samples used to measure
the PID performance has seen several improvements to face the challenges 
set by \runt of the \lhc.
The samples are now selected directly in real-time at the highest level of the 
software trigger, introducing an important benefit in terms of statistics 
and absence of selection bias with respect to the offline selection strategy
adopted in \runo.
The calibration samples are used to measure the PID performance, to 
correct the simulated samples, and to monitor the detector performance during the data-taking. 

The computing model to manage and process the calibration samples has been 
redesigned in order to overcome the scalability challenges set by the 
larger statistics needed to investigate the PID performance for the LHC
\runt. 

The new scheme has been later adopted to provide the tracking and the photon
reconstruction performance, paving the way for Run 3.




\addcontentsline{toc}{section}{References}
\setboolean{inbibliography}{true}
\bibliographystyle{LHCb}
\bibliography{main,LHCb-PAPER,LHCb-CONF,LHCb-DP,LHCb-TDR}

\newpage
 
\newpage

\end{document}